\documentclass[12pt,preprint]{aastex}

\slugcomment{Accepted for publication in ApJ}

\shorttitle{Merging Criteria of Protoplanets}
\shortauthors{Genda et al.}

\begin{document}

\title{Merging Criteria for Giant Impacts of Protoplanets}

\author{H. Genda\altaffilmark{}}
\affil{Department of Earth and Planetary Science, 
       The University of Tokyo, Hongo, Bunkyo-ku, 
       Tokyo 113-0033, Japan}
\email{genda@eps.s.u-tokyo.ac.jp}

\author{E. Kokubo\altaffilmark{}}
\affil{Division of Theoretical Astronomy, 
       National Astronomical Observatory of Japan, Osawa, Mitaka,
       Tokyo 181-8588, Japan}

\and

\author{S. Ida\altaffilmark{}}
\affil{Earth and Planetary Sciences, 
       Tokyo Institute of Technology, Ookayama, Meguro-ku, 
       Tokyo 152-8551, Japan}

\begin{abstract}

At the final stage of terrestrial planet formation, known as the giant impact stage, a few tens of Mars-sized protoplanets collide with one another to form terrestrial planets. Almost all previous studies on the orbital and accretional evolution of protoplanets in this stage have been based on the assumption of perfect accretion, where two colliding protoplanets always merge. However, recent impact simulations have shown that collisions among protoplanets are not always merging events, that is, two colliding protoplanets sometimes move apart after the collision (hit-and-run collision). As a first step towards studying the effects of such imperfect accretion of protoplanets on terrestrial planet formation, we investigated the merging criteria for collisions of rocky protoplanets. Using the smoothed particle hydrodynamic (SPH) method, we performed more than 1000 simulations of giant impacts with various parameter sets, such as the mass ratio of protoplanets, $\gamma$, the total mass of two protoplanets, $M_{\rm T}$, the impact angle, $\theta$, and the impact velocity, $v_{\rm imp}$. We investigated the critical impact velocity, $v_{\rm cr}$, at the transition between merging and hit-and-run collisions. We found that the normalized critical impact velocity, $v_{\rm cr}/v_{\rm esc}$, depends on $\gamma$ and $\theta$, but does not depend on $M_{\rm T}$, where $v_{\rm esc}$ is the two-body escape velocity. We derived a simple formula for $v_{\rm cr}/v_{\rm esc}$ as a function of $\gamma$ and $\theta$ (Eq. (\ref{eq:form_cr})), and applied it to the giant impact events obtained by \textit{N}-body calculations in the previous studies. We found that 40\% of these events should not be merging events.

\end{abstract}

%% Keywords should appear after the \end{abstract} command. The uncommented
%% example has been keyed in ApJ style. See the instructions to authors
%% for the journal to which you are submitting your paper to determine
%% what keyword punctuation is appropriate.

\keywords{accretion, accretion disk 
          --- planets and satellites: formation 
          --- solar system: formation}

%% From the front matter, we move on to the body of the paper.
%% In the first two sections, notice the use of the natbib \citep
%% and \citet commands to identify citations.  The citations are
%% tied to the reference list via symbolic KEYs. The KEY corresponds
%% to the KEY in the \bibitem in the reference list below. We have
%% chosen the first three characters of the first author's name plus
%% the last two numeral of the year of publication as our KEY for
%% each reference.

%% Authors who wish to have the most important objects in their paper
%% linked in the electronic edition to a data center may do so by tagging
%% their objects with \objectname{} or \object{}.  Each macro takes the
%% object name as its required argument. The optional, square-bracket 
%% argument should be used in cases where the data center identification
%% differs from what is to be printed in the paper.  The text appearing 
%% in curly braces is what will appear in print in the published paper. 
%% If the object name is recognized by the data centers, it will be linked
%% in the electronic edition to the object data available at the data centers  
%%
%% Note that for sources with brackets in their names, e.g. [WEG2004] 14h-090,
%% the brackets must be escaped with backslashes when used in the first
%% square-bracket argument, for instance, \object[\[WEG2004\] 14h-090]{90}).
%%  Otherwise, LaTeX will issue an error. 

\section{Introduction}

Planets are formed in a disk around a star called a protoplanetary disk, which is composed of gas and dust. Terrestrial planets are formed mainly from the dust component. Their formation process can be divided into three stages. The first stage is the formation of a large number of kilometer sized bodies called planetesimals by accretion among dust particles \citep[e.g.,][]{Goldreich73, Youdin02}. In the second stage, these planetesimals collide to produce a few tens of Mars-sized objects called protoplanets \citep[e.g.,][]{Wetherill85, Kokubo98}. The final stage is the formation of terrestrial planets from protoplanets \citep[e.g.,][]{Chambers98, Agnor99}. The collisions among protoplanets are referred to as giant impacts, and thus this final stage is known as the giant impact stage.

Giant impacts have a large influence on the various features such as the number of terrestrial planets formed, their mass and spin state \citep[e.g.,][]{Agnor99, Kokubo06}. Giant impacts are highly energetic events, and are responsible for the creation of large satellites, like the Moon \citep[e.g.,][]{Canup04a} and planets with extremely large cores such as Mercury \citep[e.g.,][]{Benz07}. Moreover, giant impacts are closely related to the thermal state such as a magma ocean \citep[e.g.,][]{Tonks92}, and the origins of the terrestrial planet atmospheres \citep{Genda05}. 

A large number of simulations of giant impacts have been devoted to the specific giant impact events related to the origin of the Moon or Mercury. However, since it is recently  believed that multiple giant impacts are common during the last stage of terrestrial planet formation, several studies \citep{Agnor04a, Asphaug09, Marcus09, Marcus10} have investigated the giant impact simulations under various impact parameters. \cite{Agnor04a} were the first to show that collisions of protoplanets during the giant impact stage are not always merging events, that is, two colliding protoplanets sometimes move apart after the collision. They called such a collision a hit-and-run collision. Except for \cite{Kokubo10}, all the previous studies on the orbital and accretional evolution of protoplanets during the giant impact stage have been based on the assumption of perfect accretion, where two colliding protoplanets always merge. However, the hit-and-run collisions demonstrated by \cite{Agnor04a} may have an important influence on many of the physical characteristics of terrestrial planets. 

In order to investigate the effects of such imperfect accretion of protoplanets on terrestrial planet formation, the merging criteria during protoplanet collisions must be clarified. \cite{Agnor04a} performed 48 simulations of the collisions between same-sized protoplanets with masses of $0.1 M_{\oplus}$, where $M_{\oplus}$ is the Earth mass. They found that hit-and-run collisions occurred when $v_{\rm imp} \ge 1.5 v_{\rm esc}$ for $\theta = 30 ^\circ$, and $v_{\rm imp} \ge 1.2 v_{\rm esc}$ for $\theta = 45^\circ$ or $60^\circ$, where $v_{\rm imp}$ is the impact velocity, $v_{\rm esc}$ is the two-body escape velocity (equation [\ref{eq:esc}]), and $\theta$ is the impact angle. Since they varied $v_{\rm imp}$ in steps of  $0.1 v_{\rm esc}$ for low-velocity collisions, the transition between the merging and hit-and-run collisions was estimated to be $v_{\rm imp} = 1.4 - 1.5 v_{\rm esc}$ for $\theta = 30 ^\circ$, and $v_{\rm imp} = 1.1 - 1.2 v_{\rm esc}$ for $\theta = 45^\circ$ or $60^\circ$. 

Subsequently, \cite{Agnor04b} investigated collisions between different-sized protoplanets with mass ratios of 1:2 and 1:10. Their results were presented in \cite{Asphaug09}. The transition was able to be estimated as follows. In the case of a 1:2 mass ratio, the transition occurs at $v_{\rm imp} = 1.4 - 1.5 v_{\rm esc}$ for $\theta = 30 ^\circ$ and $v_{\rm imp} = 1.1 - 1.2 v_{\rm esc}$ for $\theta = 45^\circ$ or $60^\circ$, whereas in a case of 1:10 mass ratio, $v_{\rm imp} = 1.5 - 2.0 v_{\rm esc}$ for $\theta = 30 ^\circ$, $v_{\rm imp} = 1.2 - 1.3 v_{\rm esc}$ for $\theta = 45^\circ$, and $v_{\rm imp} = 1.0 - 1.1 v_{\rm esc}$ for $\theta = 60^\circ$.

\cite{Marcus09} and \cite{Marcus10} investigated collisions of the rocky and icy super-Earths (up to $10 M_\oplus$), respectively. They focused on the stripping of the rocky or icy mantle resulting from a high-velocity impact. Although they did not perform the various simulations with low-impact velocities, we were able to find that the transition occurs at roughly $v_{\rm imp} = 1.0 - 1.5 v_{\rm esc}$ from their figures. 

As described above, the merging criteria has already been roughly determined for certain discrete values of the impact parameters. However, in order to carry out N-body orbital calculations with the merging criteria for the giant impact stage, a simple formula describing the dependence of the merging criteria on the impact parameters is required. To achieve this, it is necessary to determine the merging criteria over a wide range of impact parameters. In the present study, we performed more than 1000 simulations of giant impacts for various impact parameter sets using the smoothed particle hydrodynamic (SPH) method in order to formulate the merging criteria.

In Section \ref{sec:Method}, we present the SPH code and initial conditions used in our giant impact simulations. In Section \ref{sec:Outcomes}, we show the collision outcomes, and investigate the transition between merging and hit-and-run collisions. We also perform a  resolution test on the simulations. In Section \ref{sec:Criteria}, we derive the merging criteria as a function of the impact parameters, and compare the results with those of previous studies. Using the derived criteria, we then discuss the merging probability of protoplanets during the giant impact stage in Section \ref{sec:Probability}.

\section{Calculation Method}\label{sec:Method}

\subsection{Numerical Code} \label{}

In order to perform impact simulations for protoplanets, we used the SPH method \citep[e.g.,][]{Monaghan92}, which is a flexible Lagrangian method of solving hydrodynamic equations, and has been widely used in previous giant impact simulations. The SPH method can easily deal with large deformations and shock waves. Our numerical code is based on \cite{Canup04b}; here, we briefly describe its essential points. 

The equation of the motion for the \textit{i}-th SPH particle is given by
\begin{equation}
\frac{d\mathbf{v}_i}{dt} = - \sum^{\rm neighbor}_{j} \mathcal{F}_{ij} - \sum^{\rm all}_{j} \mathcal{G}_{ij},
\label{eq:SPH_EOM}
\end{equation}
where $\mathbf{v}_i$ is the velocity of the \textit{i}-th SPH particle, $t$ is the time, and $\mathcal{F}_{ij}$ and $\mathcal{G}_{ij}$ are the pressure gradient and mutual gravity terms between the \textit{i}-th and \textit{j}-th particles, respectively. Several forms have been used for the pressure gradient term, none of which appear to be clearly superior to the others. In this paper, we use the following symmetric expression,
\begin{equation}
\mathcal{F}_{ij} = m_j \left( \frac{P_i}{\rho_i^2} + \frac{P_j}{\rho_j^2} + \Pi_{ij} \right) \nabla_i W(r_{ij},h_{ij}),
\label{eq:SPH_EOM_F}
\end{equation}
where $m_j$, $P_j$, and $\rho_j$ are the mass, pressure, and density of the \textit{j}-th particle, respectively, $\Pi_{ij}$ is the artificial viscosity, $W$ is the kernel function, $r_{ij}$ is the distance between the \textit{i}-th and \textit{j}-th particles, and $h_{ij}$ is the average smoothing length of the \textit{i}-th and \textit{j}-th particles. For the artificial viscosity $\Pi_{ij}$, we use a Von Neumann-Richtmyer-type viscosity with parameters of $\alpha_{\rm vis} = 1.5$ and $\beta_{\rm vis} = 3.0$, as described in \cite{Monaghan92}. For the kernel function $W$, we use the spherically symmetric spline kernel function proposed by \citet{Monaghan85}:
\begin{equation}
W(r,h) = \frac{1}{\pi h^3} \left\{
\begin{array}{ll}
1 - \frac{3}{2} (\frac{r}{h})^2 + \frac{3}{4} (\frac{r}{h})^3, & 
0 \le \frac{r}{h} < 1, \\ 
\frac{1}{4} (2 - \frac{r}{h})^3, & 
1 \le \frac{r}{h} < 2, \\ 
0, & 
2 \le \frac{r}{h}.
\end{array}
\right.
\label{eq:SPH_kernel}
\end{equation}
This function satisfies $\int W(|\mathbf{r}|,h) d\mathbf{r} = 1$, and has a zero value when $r \ge 2 h$. In our code, $h$ is variable for each particle and with time, and determined to satisfy the condition that the number of neighboring particles ($N_{\rm nei}$) within $2h$ is  almost constant, $N_{\rm nei} = 64 \pm 2$. We used a maximum value of the smoothing length ($h_{\rm max}$) to save computational cost when searching for neighboring particles and calculating the pressure gradient term. The value of $h_{\rm max}$ is determined from
\begin{equation}
\frac{m}{\rho_{\rm min}} = \frac{4\pi}{3}(2h_{\rm max})^3,
\label{eq:hmax}
\end{equation}
where $\rho_{\rm min}$ is the minimum density, which is set to $5 \times 10^{-3}~\mathrm{kg/m^3}$ in our simulations.

The mutual gravity term in equation (\ref{eq:SPH_EOM}) can be written as 
\begin{equation}
\mathcal{G}_{ij} = G \sum_j \hat{m}_j \frac{\mathbf{r}_i-\mathbf{r}_j}{r_{ij}^3},
\label{eq:SPH_EOM_G}
\end{equation}
where $G$ is the gravitational constant, and $\hat{m}_j$ is the effective mass of the \textit{j}-th particle toward the \textit{i}-th particle defined by
\begin{equation}
\hat{m}_j = \int_0^{r_{ij}} 4 \pi r^2 m_j W(r,h_j) dr.
\label{eq:SPH_emass}
\end{equation}
This equation gives $\hat{m}_j = m_j$ when $r_{ij} \ge 2h_j$.

The mutual gravity term between all SPH particles was directly computed using a special-purpose computer for gravitational N-body systems named GRAPE-6A \citep{Fukushige05}. The GRAPE-6A can search for and produce lists of neighboring particles while simultaneously calculating their mutual gravity. This list of neighbors is used to compute the pressure gradient term in the equation of motion and the time derivative of the internal energy. In a simulation over a period of $10^5$ sec with 20,000 SPH particles, the typical CPU time was about 4 hours. Thus, the GRAPE-6A allowed us to systematically explore a wide range of impact parameters. Time integration was performed using a PEC (predict, evaluate, and correct) scheme with variable time steps \citep[e.g.,][]{Serna96}, which is second-order accurate in time.

\subsection{Pre-impact Protoplanets} \label{sec:Pre-impact}

Here we describe the modeling method for the pre-impact protoplanets. All the protoplanets are assumed to be differentiated, with a 30\% iron core and 70\% silicate mantle by mass. In our SPH simulations, we used the Tillotson equation of state \citep[]{Tillotson62}, which has been widely applied to giant impact simulations involving shock waves \citep[e.g.,][]{Benz87,Canup01,Agnor04a,Asphaug06}. The Tillotson equation of state contains ten material parameters, and the pressure is expressed as a function of the density and the specific internal energy, which is convenient for treating fluid dynamics. We used the parameter sets of granite for the silicate mantle and iron for the iron core, which are listed on page 234 of \citet{Melosh89}. 

All SPH particles in the protoplanets was set to have the same mass, and the total number of particles used for impact simulations was fixed at 20,000. For example, in the case of a collision of protoplanets with a mass ratio of 1:9, the smaller protoplanet consisted of 2,000 particles, and the larger one 18,000 particles. To model the pre-impact protoplanets, we placed the SPH particles in a 3D lattice (face-centered cubic) with iron particles on the inside, and rocky particles outside. The internal energy of the SPH particles was set to $1.0 \times 10^6$ J/kg. 

Beginning with this configuration, we calculated vibrations of the protoplanet until the particle velocities become slower than 100 m/s, which is much less than the impact velocity (order of km/s). After this operation, we used these relaxed objects as the protoplanets for impact simulation. As a first step, the protoplanets were assumed to have no spin.

In order to set impact parameters such as the impact velocity ($v_{\rm imp}$) and impact angle ($\theta$), we need to determine the radius of the pre-impact protoplanets. Since the surface boundary of the protoplanets described by the SPH particles is obscure owing to the smoothing length ($h$), we determined the radius of the pre-impact protoplanet ($R_{\rm p}$) by the following equation.
\begin{equation}
\frac{4 \pi}{3} R_{\rm p}^3 = \sum_{i} \frac{m_i}{\rho_i}.
\label{eq:radius}
\end{equation}

\subsection{Initial Conditions for Collisions} \label{subsec:IC}

We prepared more than 1000 sets of initial conditions for the giant impact simulations. The parameters used were the mass ratio of the protoplanets ($\gamma = M_{\rm i}/M_{\rm t}$, where $M_{\rm t}$ and $M_{\rm i}$ are the mass of the target and impactor, respectively), the total mass of the two protoplanets ($M_{\rm T} = M_{\rm i} + M_{\rm t}$), the impact angle ($\theta$), and the impact velocity ($v_{\rm imp}$). We systematically varied the mass ratio as $\gamma = $ 1, 2/3, 1/2, 1/3, 1/4, 1/6, and 1/9. For $\gamma = $ 1, 1/4, and 1/9, we considered three different values of $M_{\rm T}$. In total, we used 13 different combinations for the two colliding protoplanets (see table~\ref{mass}). For each mass combination, we varied the impact angle in the range $\theta = 0^\circ-75^\circ$ in $15^\circ$ steps, and the impact velocity in the range $v_{\rm imp} = 1.0-3.0 v_{\rm esc}$ in $0.2 v_{\rm esc}$ steps, where $v_{\rm esc}$ is the two-body escape velocity defined as
\begin{equation}
v_{\rm esc} = \sqrt{\frac{2 G M_{\rm T}}{R_{\rm t} + R_{\rm i}}},
\label{eq:esc}
\end{equation}
where $R_{\rm t}$ and $R_{\rm i}$ are the radius of the target (larger protoplanet) and the impactor (smaller protoplanet), respectively. To precisely determine the transition between merging and hit-and-run collisions, we varied $v_{\rm imp}$ with a smaller step size of 0.02 $v_{\rm esc}$ near the transition. In total, we performed more than 1000 runs, consisting of 13 (mass combinations) $\times$ 6 (angles) $\times$ 16 (11 runs with 0.2$v_{\rm esc}$ steps and $\sim$ 5 runs with 0.02$v_{\rm esc}$ steps).

\begin{figure}
\epsscale{.5}
\plotone{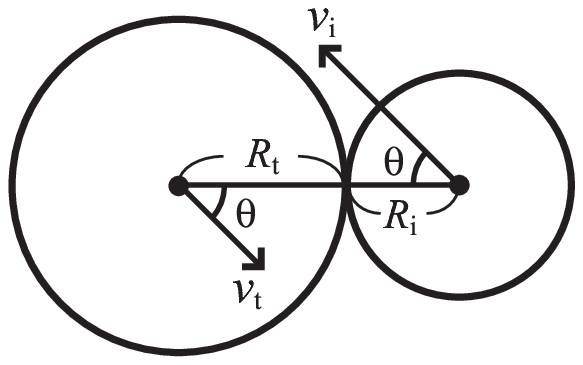}
\caption{Geometry of the collision between a larger (target) and smaller (impactor) protoplanets with radii of $R_{\rm t}$ and $R_{\rm i}$, respectively. Since a center of mass coordinate system is used, the impact velocity, $v_{\rm imp}$, is given by $|v_{\rm i} - v_{\rm t}|$, where $v_{\rm t}$ and $v_{\rm i}$ are the velocities of the larger and smaller protoplanets, respectively. The impact angle is $\theta$. A head-on collision corresponds to $\theta = 0^\circ$.}
\label{fig:impact_angle}
\end{figure}

The impact parameters $v_{\rm imp}$ and $\theta$ are defined when the two protoplanets are in contact with each other (see Figure \ref{fig:impact_angle}). We assumed that the two protoplanets are mass points, and calculated backward the positions of two mass points until their distance was apart at a distance of 3 ($R_{\rm i}$+$R_{\rm t}$). Then, we performed the giant impact simulations over a period of $10^5$ sec.

\section{Collision Outcomes}\label{sec:Outcomes}

\subsection{Merging and Hit-and-Run Collisions} \label{}

The outcomes of the collisions between the protoplanets are divided into two types: merging and hit-and-run collisions. According to the previous studies \citep[]{Asphaug10, Leinhardt10, Leinhardt11}, the collision outcomes are subdivided into several regimes (e.g., partial accretion, fragmentation and so on). However, we classify collision outcomes into only two regimes of merging and hit-and-run collisions, because it is the most essential to the evolution of protoplanets during the giant impact stage that only one big body is left after the giant impact (i.e., merging collision) or two big bodies are left (i.e., hit-and-run collision). Additionally, we need to classify collision outcomes as simply as possible, in order to incorporate those into \textit{N}-body simulation. 

\begin{figure}
\epsscale{0.75}
\plotone{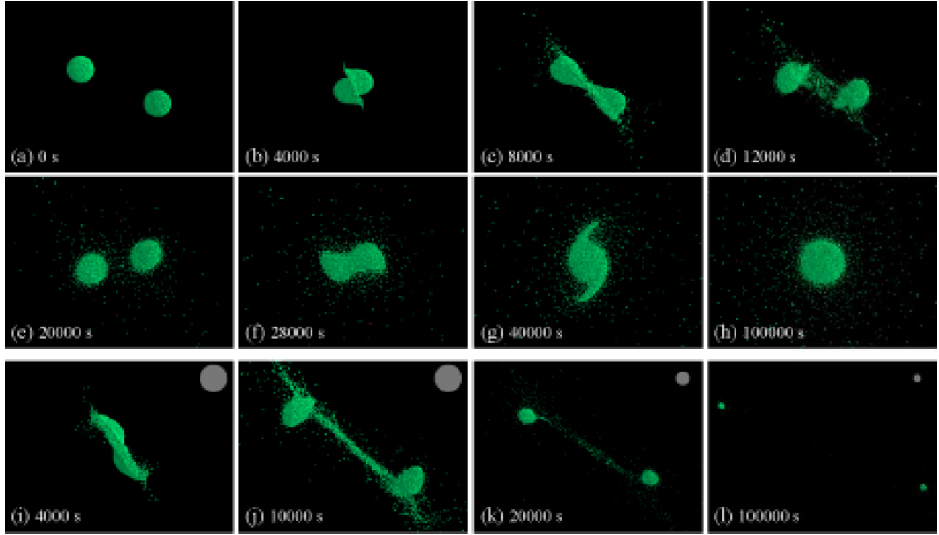}
\caption{Snapshots of two typical giant impacts between equal-mass protoplanets with $0.1 M_{\oplus}$.  Panels (a) to (h) show the time sequence for a relatively low-velocity collision of protoplanets ($v_{\rm imp} = 1.3 v_{\rm esc}$) with the impact angle $30^\circ$. The colliding protoplanets finally merge so that this type of collision is referred to as a merging collision. Panels (i) to (l) show the time sequence for a relatively high-velocity collision ($v_{\rm imp} = 1.5 v_{\rm esc}$). The impact angle and mass of the protoplanets are the same as in (a) to (h), but the protoplanets do not merge. This type of collision is referred to as a hit-and-run collision. The dark gray circle at the top-right corner in panels (i) to (l) indicates the size of the initial protoplanets.}
\label{fig:snapshots}
\end{figure}

Figure \ref{fig:snapshots} shows snapshots of two typical collisions. Panels (a) to (h) in Figure \ref{fig:snapshots} show the time sequence for a relatively low-velocity collision ($v_{\rm imp} = 1.3 v_{\rm esc}$) of same-sized protoplanets ($M_{\rm i} = M_{\rm t} = 0.1 M_{\oplus}$) with $\theta = 30 ^\circ$. After the first contact, the protoplanets become separated (see panel (e)), but remain gravitationally bound. Although some amount of mantle material is ejected, almost all parts of the colliding protoplanets finally merge. We refer to this type of collision as ``a merging collision''. On the other hand, a relatively high-velocity collision leads to a completely different result. Panels (i) to (l) show the time sequence of a collision with $v_{\rm imp} = 1.5 v_{\rm esc}$. The impact angle and protoplanet masses are the same as in panels (a) to (h). After the first contact, the protoplanets escape from each other and are no longer gravitationally bound. We refer to this type of collision as ``a hit-and-run collision''. 
By examining snapshots at $t = 10^5 \mathrm{sec}$, we could easily classify almost all the simulations performed in this study as either merging or hit-and-run collisions. However, in some cases, the protoplanets were separated but still gravitationally bound at $t = 10^5~\mathrm{sec}$. Since those protoplanets are expected to eventually merge, we classified such a case as a merging collision.

\subsection{Mass of Protoplanet after a Collision} \label{}

Here, we consider the mass of the protoplanet after the collision. We briefly describe the method to determine the mass of the gravitationally bound objects from the SPH particle data after $t = 10^5~\mathrm{sec}$. In the first step, we roughly identify clumps of SPH particles using a friends-of-friends algorithm \citep[e.g.,][]{Huchra82}. We then iteratively check whether or not any SPH particles not belonging to clumps are gravitationally bound to clumps. Finally, we iteratively identify pairs of clumps that are gravitationally bound. Such clumps should merge after $t = 10^5~\mathrm{sec}$, and we can thus regard them as a single object. We define the mass of the largest object as $M_1$ and the second largest object as $M_2$.

Figure \ref{fig:result_1vs1_30deg} shows the mass of the largest object normalized by the total mass, $M_1/M_{\rm T}$, as a function of the impact velocity normalized by the escape velocity, $v_{\rm imp}/v_{\rm esc}$, for collisions of same-sized protoplanets with $\theta = 30 ^\circ$. This figure also shows the numerical results of \cite{Agnor04a} for collisions of same-sized protoplanets with $0.1 M_{\oplus}$, which are in good agreement with our results for total mass $M_{\rm_T}$ of $0.2 M_{\oplus}$. We also performed simulations for $M_{\rm T} = 0.4 M_{\oplus}$ and $1.0 M_{\oplus}$. As seen in Figure \ref{fig:result_1vs1_30deg}, $M_1/M_{\rm T}$ does not depend on the total mass, when the normalized impact velocity, $v_{\rm imp}/v_{\rm esc}$, is considered. The insensitivity to the total mass holds true to collisions between protoplanets, which has been predicted by \citet{Asphaug10}. If the material properties such as strength dominates over gravity (typically collision between smaller bodies with less than 1 km in radius), the collision outcomes would depend on the total mass \citep[]{Asphaug10}.

\subsection{Transition between Merging and Hit-and-Run Collisions} \label{}

\begin{figure}
\epsscale{1.0}
\plotone{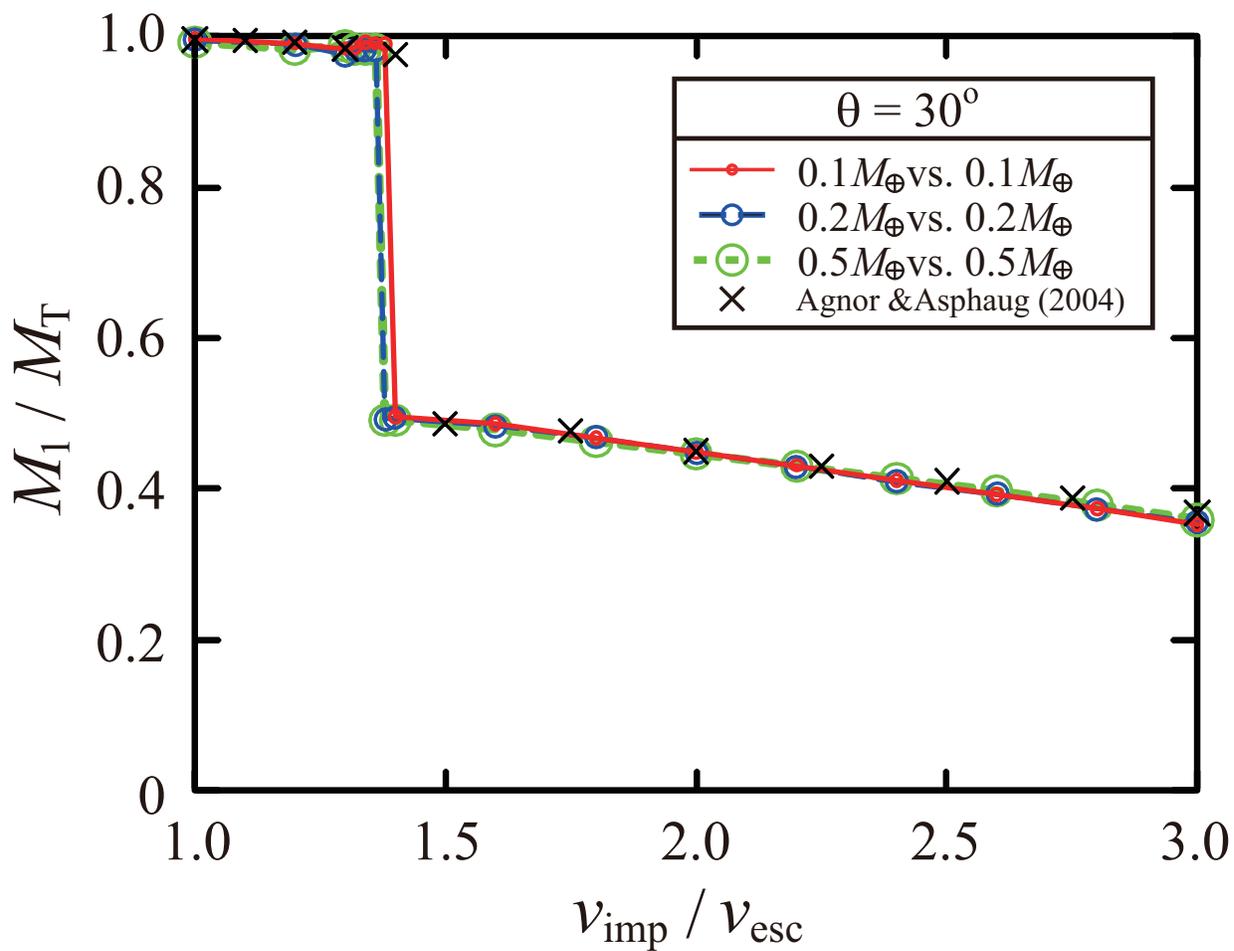}
\caption{Normalized mass of the largest gravitationally bound object, $M_1/M_{\rm T}$, as a function of normalized impact velocity, $v_{\rm imp}/v_{\rm esc}$. Data for $\theta = 30^{\circ}$ impacts between two equal-mass protoplanets with different total masses are plotted. Our collision outcomes are very similar to those obtained by \cite{Agnor04a} who performed simulations for collisions of same-sized protoplanets with masses of $0.1M_{\oplus}$. }
\label{fig:result_1vs1_30deg}
\end{figure}

Figure \ref{fig:result_1vs1_30deg} indicates that $M_1/M_{\rm T}$ changes sharply around $1.4 v_{\rm esc}$. Collisions at impact velocities less than this velocity result in almost perfect accretion (i.e., $M_1/M_{\rm T} \simeq 1$), and thus are classified as merging collisions. For the case of a near head-on collision ($\theta \le 15 ^\circ$), $M_1/M_{\rm T}$ gradually decreases with the increase of $v_{\rm imp}$. Although such a collision should be classified as a partial accretion collision or a fragmentation collision according to \cite{Asphaug10} and \cite{Leinhardt11}, we here classify those as a merging collision because only one large body remains after the collision. Since the probability of a near head-on collision with high velocity is quite low during the giant impact stage, our treatment would not become a serious problem. On the other hand, Figure \ref{fig:result_1vs1_30deg} indicates that the collisions at impact velocities higher than $1.4 v_{\rm esc}$ result in $M_1/M_{\rm T} \simeq 0.5$. Some amount of material become stripped from the protoplanets, and the protoplanets escape from each other. These collisions are classified as hit-and-run collisions. We refer to the impact velocity at the transition between merging and hit-and-run collisions as the critical impact velocity, $v_{\rm cr}$. The normalized critical impact velocity, $v_{\rm cr}/v_{\rm esc}$, is not strongly dependent on the total mass, $M_{\rm T}$. We obtain $v_{\rm cr}/v_{\rm esc}$ = $1.39 \pm 0.01$, $1.37 \pm 0.01$, and $1.37 \pm 0.01$ for $M_{\rm T} = 0.2, 0.4$ and $1.0 M_{\oplus}$, respectively.

\begin{figure}
\epsscale{1.0}
\plotone{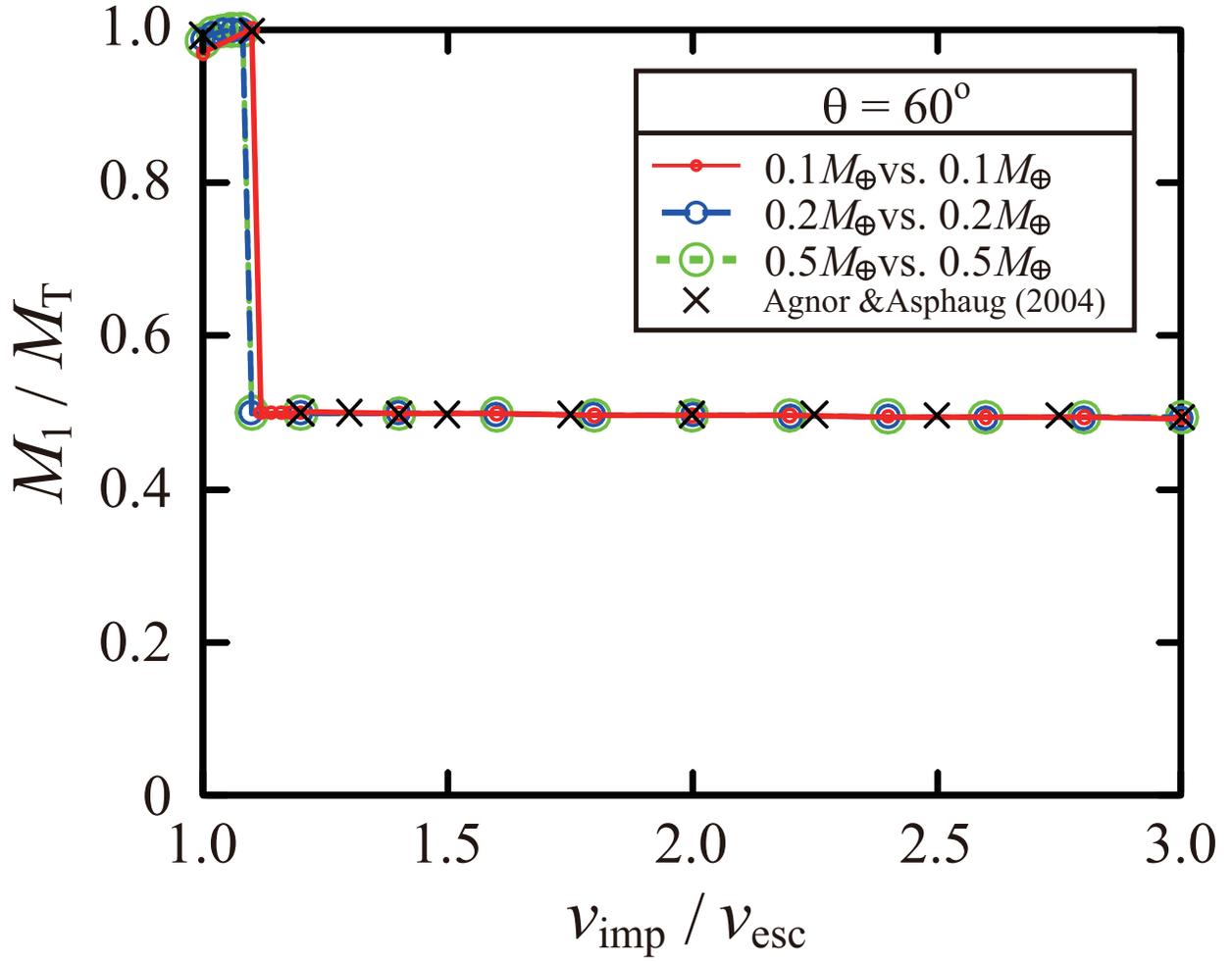}
\caption{The same as Figure \ref{fig:result_1vs1_30deg} but for $\theta = 60^{\circ}$.}
\label{fig:result_1vs1_60deg}
\end{figure}

The critical impact velocity is expected to depend on the impact angle. Figure \ref{fig:result_1vs1_60deg} is similar to Figure \ref{fig:result_1vs1_30deg}, but for the collisions with $\theta = 60 ^\circ$. As is the same in the case of $\theta = 30 ^\circ$, $M_1/M_{\rm T}$ for $\theta = 60 ^\circ$ does not depend on the total mass. However, collisions with $\theta = 60 ^\circ$ result in lower $v_{\rm cr}/v_{\rm esc}$ values than the $\theta = 30 ^\circ$ case. The calculated values are $v_{\rm cr}/v_{\rm esc}$ = $1.11 \pm 0.01$, $1.09 \pm 0.01$, and $1.09 \pm 0.01$ for $M_{\rm T} = 0.2, 0.4$ and $1.0 M_{\oplus}$, respectively. This result implies that collisions at higher impact angles are more like to be hit-and-run collisions. 

\begin{figure}
\epsscale{1.0}
\plotone{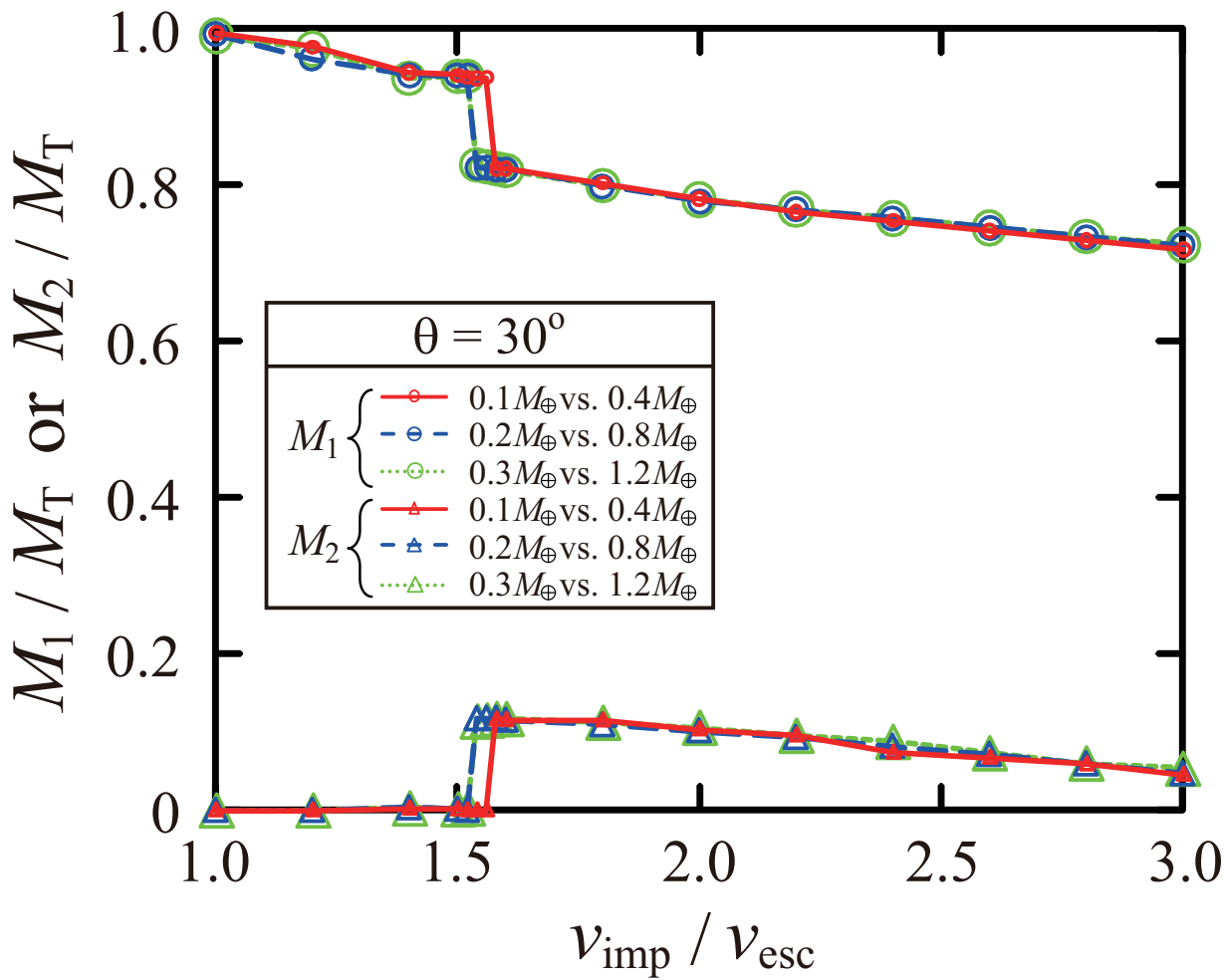}
\caption{The same as Figure \ref{fig:result_1vs1_30deg} but for collisions of different-sized protoplanets with the mass ratio of 1:4 ($\gamma = 1/4$). The mass of the second-largest object, $M_2$, is also plotted.}
\label{fig:result_1vs4_30deg}
\end{figure}

The critical impact velocity is also expected to depend on the mass ratio of the protoplanets ($\gamma$). Figure \ref{fig:result_1vs4_30deg} shows the results for $\gamma = 1/4$ (mass ratio of 1:4) and $\theta = 30 ^\circ$. It can be seen that in this case also, $M_1/M_{\rm T}$ does not depend on the total mass. This has also been verified for $\gamma = 1/9$ (mass ratio of 1:9), although the results are not shown here. As shown in Figure \ref{fig:result_1vs4_30deg}, $M_1/M_{\rm T}$ changes sharply around $1.5 v_{\rm esc}$. The normalized critical impact velocities are $v_{\rm cr}/v_{\rm esc}$ = $1.57 \pm 0.01$, $1.53 \pm 0.01$, and $1.53 \pm 0.01$ for $M_{\rm T} = 0.5, 1.0$ and $1.5 M_{\oplus}$, respectively, which are larger than for the case of collisions between same-sized protoplanets.

\subsection{Dependence on Particle Number and Initial Internal Energy} \label{}

In addition to the impact simulations with 20,000 SPH particles (standard case), we performed simulations with 3,000 (low-resolution case), 60,000 and 100,000 particles (high-resolution cases) for certain impact parameters in order to check the dependence of convergence on particle number. Although we found that the critical impact velocity for the low-resolution case was slightly different from that for the standard case, the high-resolution cases yielded the same results. For example, for $\gamma = 1$, $M_{\rm T} = 0.2 M_\oplus$, and $\theta = 30^\circ$, we obtained $v_{\rm cr}/v_{\rm esc} = 1.39 \pm 0.01$ for both the standard and high-resolution cases. Therefore, using 20,000 SPH particles is enough for determining the critical impact velocity.

In subsection \ref{sec:Pre-impact}, the initial internal energy of the SPH particles was set to $1.0 \times 10^6$ J/kg. To investigate the effect of the initial thermal state of the protoplanets, we prepared pre-impact protoplanets with an internal energies of $1.0 \times 10^4$ (cold-state case) and $3.0 \times 10^6$ J/kg (hot-state case). We then performed simulations of collisions between same-sized protoplanets with impact parameters near the transition between merging and hit-and-run collisions. We found that although the results for the very low-impact angle case ($\theta = 15^\circ$) showed a very slight dependence on the initial thermal state, for all other cases no such dependence was observed.

\section{Merging Criteria for Colliding Protoplanets}\label{sec:Criteria}

\begin{figure}
\epsscale{1.0}
\plotone{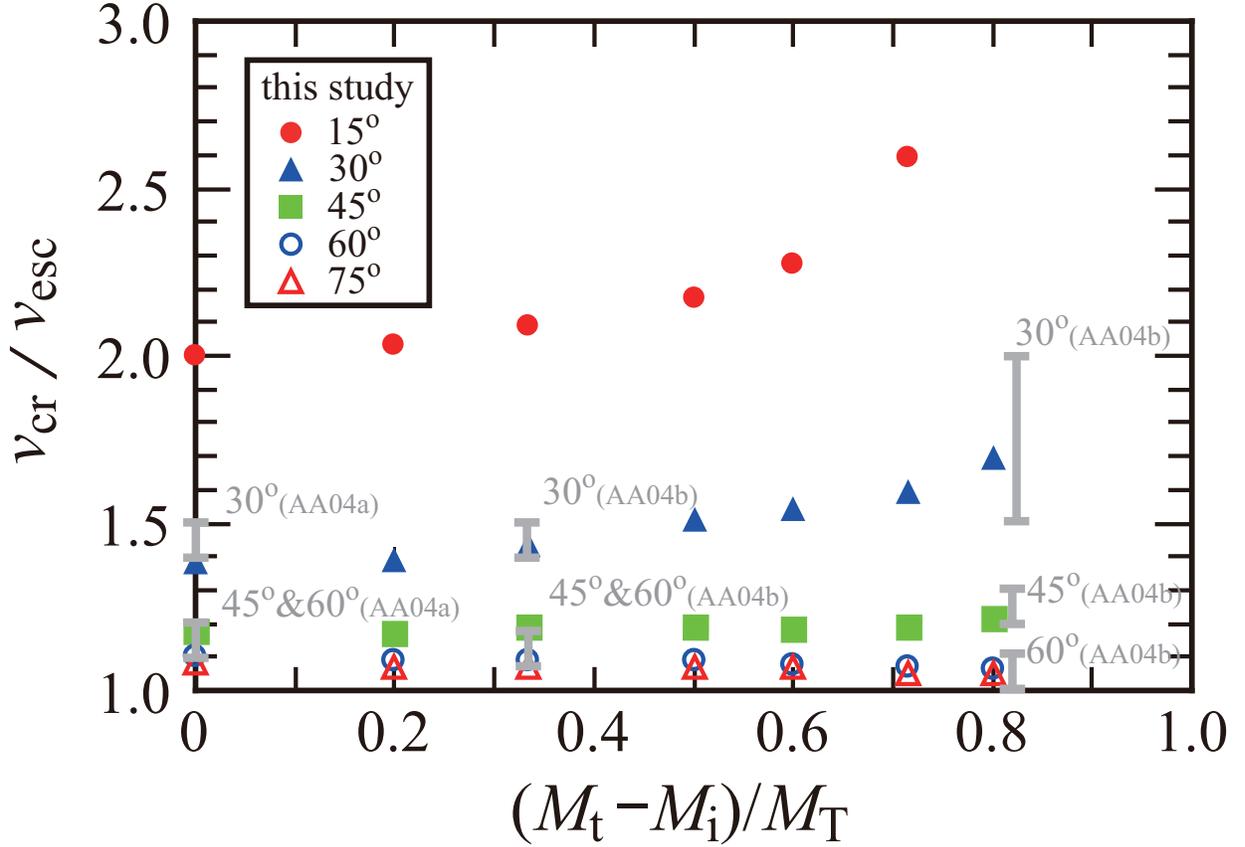}
\caption{Critical impact velocities for the cases of the various impact angles and mass ratios of protoplanets. Note that $(M_{\rm t} - M_{\rm i})/M_{\rm T}$ is a function of $\gamma$, i.e., $(1 - \gamma)/(1 + \gamma)$. Data points represent our numerical results for $\theta = 15^\circ$ (filled circles), $30^\circ$ (filled triangles), $45^\circ$ (filled squares), $60^\circ$ (open circles), and $75^\circ$ (open triangles). Bars labeled with AA04a and AA04b are the results obtained by \cite{Agnor04a} and \cite{Agnor04b}, respectively.}
\label{fig:cr_data}
\end{figure}

In the previous section, we determined the critical impact velocities for several impact parameters. In Figure \ref{fig:cr_data}, we summarize $v_{\rm cr}/v_{\rm esc}$ for all parameter sets of the mass ratios ($\gamma$) and impact angles ($\theta$). For $\gamma = 1$, $1/4$, and $1/9$, we performed simulations for three sets of $M_{\rm T}$ (see Table \ref{mass}). In Figure \ref{fig:cr_data}, we plot the average of these three results. 

The critical impact velocities estimated by \cite{Agnor04a} and \cite{Agnor04b} are also plotted in Figure \ref{fig:cr_data}. \cite{Agnor04a} investigated collisions between same-sized protoplanets with masses of $0.1 M_{\oplus}$, and found that $v_{\rm cr}/v_{\rm esc} = 1.4 - 1.5$ for $\theta = 30 ^\circ$, and $v_{\rm cr}/v_{\rm esc} = 1.1 - 1.2$ for $\theta = 45^\circ$ or $60^\circ$. Our results are in good agreement with those of \cite{Agnor04a}, since in our simulations, $v_{\rm cr}/v_{\rm esc} = 1.39 \pm 0.01$ for $\theta = 30 ^\circ$, $1.19 \pm 0.01$ for $\theta = 45^\circ$, and $1.11 \pm 0.01$ for $\theta = 60^\circ$.

\cite{Agnor04b} investigated collisions between different-sized protoplanets with mass ratios of 1:2 and 1:10. Based on Figure 17 in \cite{Asphaug09}, for a mass ratio of 1:2, $v_{\rm cr}/v_{\rm esc} = 1.4 - 1.5$ for $\theta = 30 ^\circ$, and $1.1 - 1.2$ for $\theta = 45 ^\circ$ or $60 ^\circ$. In the present study, for a mass ratio of 1:2, we obtained $v_{\rm cr}/v_{\rm esc} = 1.43 \pm 0.01$ for $\theta = 30 ^\circ$, $1.19 \pm 0.01$ for $\theta = 45^\circ$, and $1.09 \pm 0.01$ for $\theta = 60^\circ$, and these values are in good agreement with those of \cite{Agnor04b}. Although we did not perform simulations for a mass ratio of 1:10, our data points for 1:9 ($\gamma = 1/9$) seem to fall within the range obtained by \cite{Agnor04b}. 
 
As shown in Figure \ref{fig:cr_data}, $v_{\rm cr}/v_{\rm esc}$ increases with decreasing impact angle or mass ratio, which means that collisions with low impact angles or low mass ratios tend to be merging events. This can be explained in terms of the size of the overlapping volume of the colliding protoplanets. Since this is geometrically smaller for higher impact angle, the fraction of kinetic energy converted to thermal energy of protoplanets and kinetic energy of the fragments is small, resulting in a hit-and-run collision. In addition, in the case of a small impactor (i.e., small $\gamma$), most of the volume of the impactor tends to overlap with the target. Therefore, the impactor can not easily be ejected, which leads to be a merging collision.

\begin{figure}
\epsscale{1.0}
\plotone{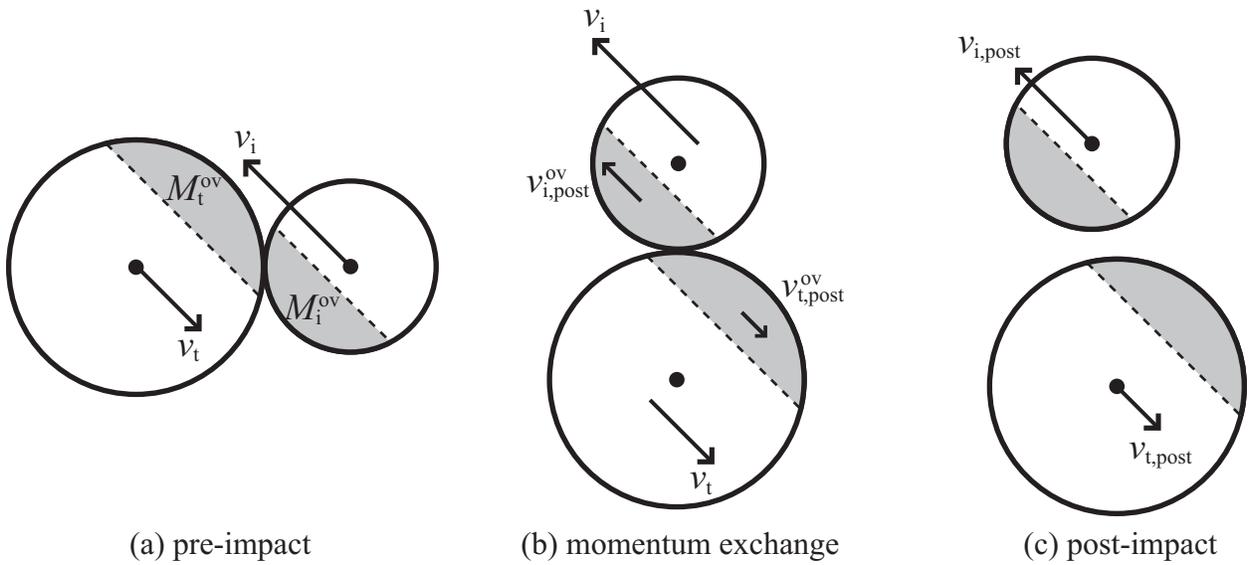}
\caption{Configuration of a target and an impactor for the simple physical model. (a) Before the impact, the velocities of the target and the impactor are $v_{\rm t}$ and $v_{\rm i}$, respectively. (b) During the impact, the overlapping parts (shaded areas) exchange momentum, and the post-impact velocity of the overlapping area is reduced to $v_{\rm t,post}^{\rm ov}$ for the target and $v_{\rm i,post}^{\rm ov}$ for the impactor. (c) After the impact, velocities of the target and impactor are $v_{\rm t,post}$ and $v_{\rm i,post}$, respectively.}
\label{fig:simple_model}
\end{figure}

In the following, we consider a simple physical model in order to express the critical impact velocity as a function of the impact angle and mass ratio. For two spheres with radii $R_{\rm t}$ and $R_{\rm i}$ colliding with an impact angle $\theta$ (see Figure \ref{fig:simple_model}), the mass fractions of the overlapping volumes for the target and impactor ($\beta_{\rm t}$ and $\beta_{\rm i}$, respectively) are geometrically given by
\begin{equation}
\beta_{\rm t} = \frac{M_{\rm t}^{\rm ov}}{M_{\rm t}} = \biggl( \frac{R_{\rm t}+R_{\rm i}}{R_{\rm t}} \biggr)^2 (1-\sin\theta)^2 \Biggl\{ \frac{3}{4} - \frac{1}{4} \biggl( \frac{R_{\rm t}+R_{\rm i}}{R_{\rm t}} \biggr) (1-\sin\theta) \Biggr\},
\label{eq:overlap_target}
\end{equation}
\begin{equation}
\beta_{\rm i} = \frac{M_{\rm i}^{\rm ov}}{M_{\rm i}} = \left\{
\begin{array}{cl}
\biggl( \frac{R_{\rm t}+R_{\rm i}}{R_{\rm i}} \biggr)^2 (1-\sin\theta)^2 \Biggl\{ \frac{3}{4} - \frac{1}{4} \biggl( \frac{R_{\rm t}+R_{\rm i}}{R_{\rm i}} \biggr) (1-\sin\theta) \Biggr\}, &
\rm if ~\sin\theta \ge 1-\frac{2R_{\rm t}}{R_{\rm i} + R_{\rm t}}, \\
1, &
\rm otherwise,
\end{array}
\right.
\label{eq:overlap_impactor}
\end{equation}
where $M_{\rm t}^{\rm ov}$ and $M_{\rm i}^{\rm ov}$ are the masses of the overlapping volumes for the target and impactor, respectively. A constant density is assumed for simplicity.

As illustrated in Figure \ref{fig:simple_model}, we divide the spheres into overlapping parts and non-overlapping parts, and consider the momentum exchange between the overlapping parts. Using the parameter of the degree of the momentum exchange ($\alpha$), the post impact velocities of the overlapping parts ($v_{\rm t,post}^{\rm ov}$ and $v_{\rm i,post}^{\rm ov}$) are expressed as
\begin{equation}
\left\{
\begin{array}{l}
v_{\rm t,post}^{\rm ov} = (1-\alpha)(v_{\rm t}-v_{\rm COM}^{\rm ov}) + v_{\rm COM}^{\rm ov}, \\
v_{\rm i,post}^{\rm ov} = (1-\alpha)(v_{\rm i}-v_{\rm COM}^{\rm ov}) + v_{\rm COM}^{\rm ov},
\end{array}
\right.
\label{eq:overlap_post_velocity}
\end{equation}
where $v_{\rm COM}^{\rm ov}$ is the velocity of the center of mass of the overlapping parts, and is written as
\begin{equation}
v_{\rm COM}^{\rm ov} = \frac{M_{\rm t}^{\rm ov} v_{\rm t} + M_{\rm i}^{\rm ov} v_{\rm i}}{M_{\rm t}^{\rm ov} + M_{\rm i}^{\rm ov}}.
\end{equation}
For example, when $\alpha = 1$, the velocities of the overlapping parts ($v_{\rm t,post}^{\rm ov}$ and $v_{\rm i,post}^{\rm ov}$) become $v_{\rm COM}^{\rm ov}$ owing to the complete momentum exchange. 

\begin{figure}
\epsscale{1.0}
\plotone{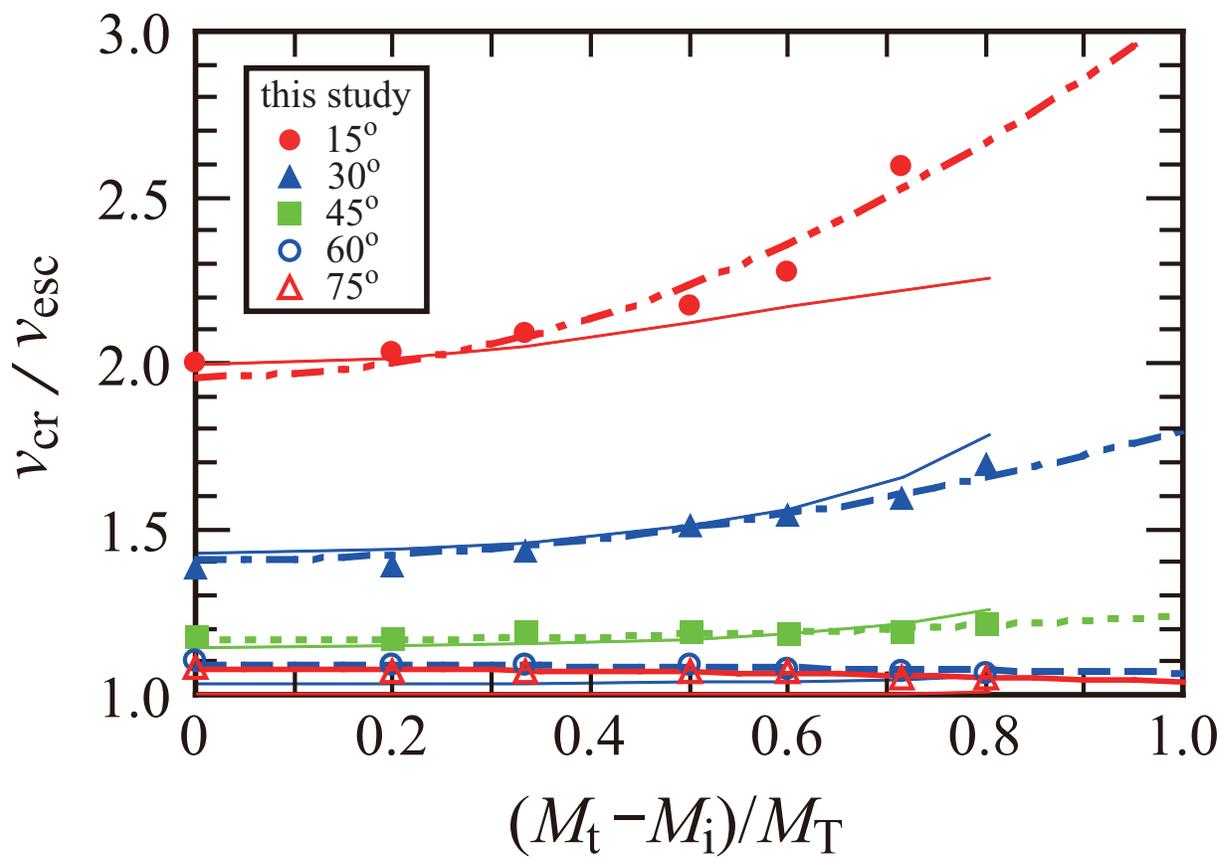}
\caption{Fits to the normalized critical impact velocity data. Data points are our numerical results and the same as shown in Figure \ref{fig:result_1vs1_30deg}. Thin and thick curves represent fits using equation (\ref{eq:form_cr_phys}) with $\alpha = 0.6$, and equation (\ref{eq:form_cr}), respectively.}
\label{fig:cr_fit}
\end{figure}

The post impact velocities of the entire target and impactor can be derived based on the conservation of momentum, and written as
\begin{equation}
\left\{
\begin{array}{l}
M_{\rm t} v_{\rm t,post} = (M_{\rm_t}-M_{\rm t}^{\rm ov}) v_{\rm t} + M_{\rm t}^{\rm ov} v_{\rm t,post}^{\rm ov}, \\
M_{\rm i} v_{\rm i,post} = (M_{\rm_i}-M_{\rm i}^{\rm ov}) v_{\rm i} + M_{\rm i}^{\rm ov} v_{\rm i,post}^{\rm ov},
\end{array}
\right.
\label{eq:post_mass_cons}
\end{equation}
where $v_{\rm t,post}$ and $v_{\rm i,post}$ are the velocities for the post-impact target and impactor, respectivery. Combining equations. (\ref{eq:overlap_post_velocity}) - (\ref{eq:post_mass_cons}) gives  
\begin{equation}
\left\{
\begin{array}{l}
v_{\rm t,post} = \biggl( 1 - \alpha \beta_{\rm t} \beta_{\rm i} \frac{M_{\rm t} + M_{\rm i}}{\beta_{\rm t} M_{\rm t} + \beta_{\rm i} M_{\rm i}} \biggr) v_{\rm t}, \\
v_{\rm i,post} = \biggl( 1 - \alpha \beta_{\rm t} \beta_{\rm i} \frac{M_{\rm t} + M_{\rm i}}{\beta_{\rm t} M_{\rm t} + \beta_{\rm i} M_{\rm i}} \biggr) v_{\rm i}.
\end{array}
\right.
\end{equation}
If the relative velocity of post-impact objects is higher than the two-body escape velocity, a hit-and-run collision should occur. Therefore, to solve $v_{\rm i,post} - v_{\rm t,post} = v_{\rm esc}$ using $v_{\rm cr} = v_{\rm i} - v_{\rm t}$, the critical impact velocity is obtained as follows:
\begin{equation}
\frac{v_{\rm cr}}{v_{\rm esc}}= \Biggl[ 1 - \alpha \beta_{\rm t} \beta_{\rm i} \frac{M_{\rm t} + M_{\rm i}}{\beta_{\rm t} M_{\rm t} + \beta_{\rm i} M_{\rm i}} \Biggr]^{-1}.
\label{eq:form_cr_phys}
\end{equation}
The calculated results for $\alpha = 0.6$ are drawn as gray curves in Figure \ref{fig:cr_fit}. We could roughly reproduce the dependence of $v_{\rm cr}/v_{\rm esc}$ on $\theta$ and $\gamma$, but detail features, especially the case for the low-impact angle and low-mass ratio, could not be reproduced. For such a collision, the role of fragmentation that has not been considered here may become important. It should be also noted that $\alpha = 0.6$ would not be applied to collisions other than giant impacts between rocky protoplanets.  

In addition to the above physical model, we tried to mathematically fit the numerical data for the critical impact velocity. Of the many possibilities available, we found that the following simple formula with five fitting parameters, $c_1$ to $c_5$, was most effective, 
\begin{equation}
\frac{v_{\rm cr}}{v_{\rm esc}}=c_1 \mit \Gamma \Theta^{c_5} +c_2\Gamma+c_3 \Theta^{c_5}+c_4,
\label{eq:form_cr}
\end{equation}
where $\mathit\Gamma = (1-\gamma)/(1+\gamma) = (M_{\rm t}-M_{\rm i})/M_{\rm T}$, and $\mathit\Theta = 1-\sin\theta$. The fitting parameters are $c_1 = 2.43$, $c_2 = -0.0408$, $c_3 = 1.86$, $c_4 = 1.08$, and $c_5 = 5/2$. The fitting curves produced by equation (\ref{eq:form_cr}) are shown in Figure \ref{fig:cr_fit} as thick curves, and are in excellent agreement with numerical results. The value of the fitting parameters derived here is limited to the collision between protoplanets.

\section{Merging Probability of Protoplanets}\label{sec:Probability}

Almost all previous N-body simulations of terrestrial planet formation during the giant impact stage have been performed based on the assumption of perfect accretion. In order to investigate the statistical properties of fully formed terrestrial planets, \citet{Kokubo06} considered 10 sets of protoplanet initial conditions, and performed 20 runs for each set under the assumption of perfect accretion. Subsequently, to investigate the spin state of the formed planets, \citet{Kokubo07} performed additional 30 runs each for 7 sets of protoplanet initial conditions. Using the formula for the merging criteria (equation [\ref{eq:form_cr}]) derived in the present study, we can now determine whether each of giant impacts was a merging or a hit-and-run event.

\begin{figure}
\epsscale{1.0}
\plotone{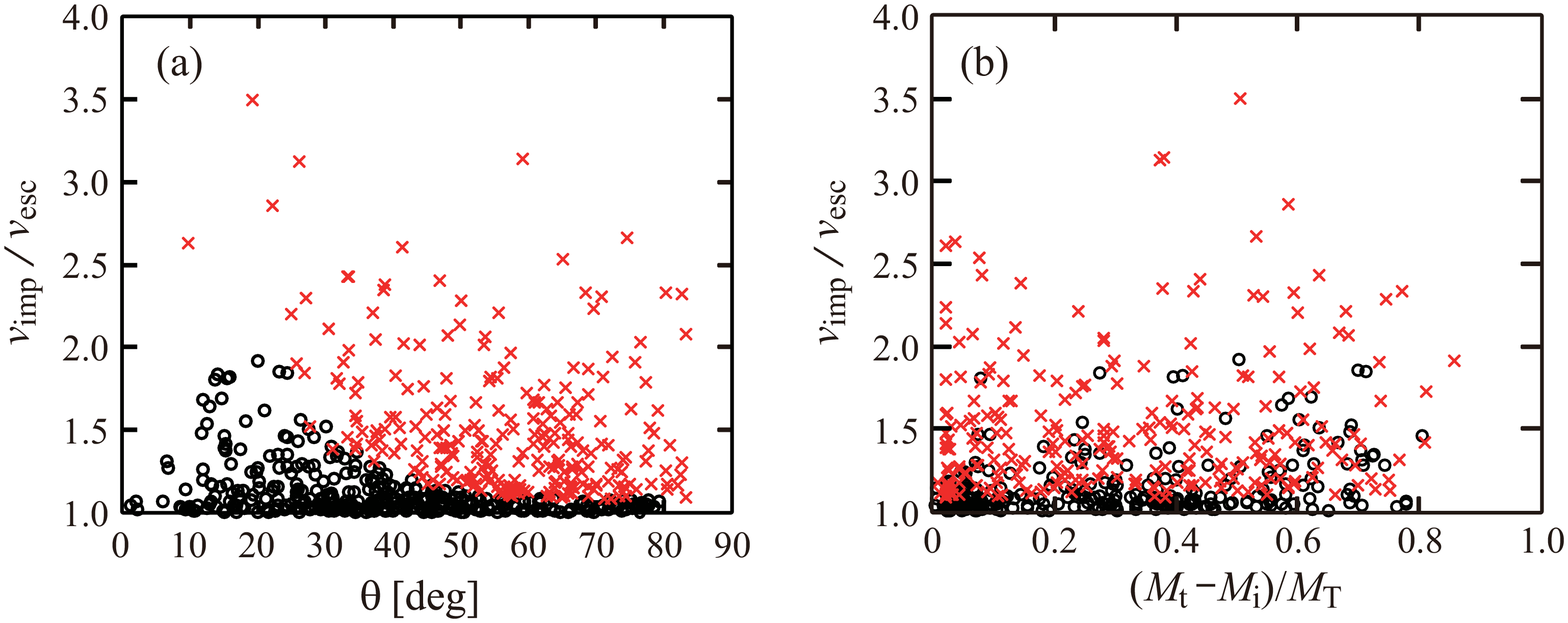}
\caption{Normalized impact velocity for 635 giant impacts reported by \cite{Kokubo06} and \cite{Kokubo07} as a function of the impact angle (a) and the mass ratio of protoplanets (b). Based on the critical impact velocity (equation [\ref{eq:form_cr}]), the giant impact events are distinguished as hit-and-run collisions (cross symbols) or merging collisions (circle symbols). Although \cite{Kokubo06} and \cite{Kokubo07} assumed the perfect accretion of protoplanets in their N-body calculations, the present study reveals that 40\% of the impact events (256 out of 635) are hit-and-run collisions.}
\label{fig:impact_events}
\end{figure}

Figure \ref{fig:impact_events} shows the normalized impact velocity as a function of impact angle (left panel) and mass ratio (right panel) for 635 giant impact events during 50 runs under standard initial conditions (Model 1 in \citet{Kokubo07}). The symbols denoted by crosses represent hit-and-run events, as determined by equation (\ref{eq:form_cr}). In fact, 40\% of all impact events (256 out of 635) are expected to be hit-and-run collisions. This result is consistent with a previous study by \cite{Agnor04a}, who estimated a hit-and-run probability of roughly half. 

\section{Summary and Discussion}

During the giant impact stage of terrestrial planet formation in our solar system, a few tens of Mars-sized protoplanets collide with each other to form terrestrial planets. Almost all previous studies on N-body calculations of the giant impact stage have been based on the assumption of perfect accretion. However, recent impact simulations have shown that collisions of protoplanets are not always merging events. 

As a first step towards studying the effects of such imperfect accretion on terrestrial planet formation, we investigated the merging criteria for a collision of rocky protoplanets. Using the SPH method, we performed more than 1000 simulations of giant impacts for various parameter sets, such as the mass ratio of colliding protoplanets ($\gamma$), the total mass of two protoplanets ($M_{\rm T}$), the impact angle ($\theta$), and the impact velocity ($v_{\rm imp}$). We investigated the critical impact velocity ($v_{\rm cr}$) at the transition between merging and hit-and-run collisions. We found that the normalized critical impact velocity, $v_{\rm cr}/v_{\rm esc}$, depends on $\gamma$ and $\theta$, but does not depend on $M_{\rm T}$. We derived a simple formula for $v_{\rm cr}/v_{\rm esc}$ as a function of $\gamma$ and $\theta$ (see equation [\ref{eq:form_cr}]), and applied it to the giant impact events considered by \citet{Kokubo06} and \citet{Kokubo07}. We found that 40\% of these events should not be merging events.

\cite{Kokubo10} was the first to performed N-body simulations of the giant impact stage taking into account the merging criteria shown in equation (\ref{eq:form_cr}), and investigated the effects of imperfect accretion on terrestrial planet formation. They found that some basic properties such as the final number, mass, orbital elements, and growth timescale of planets did not change very much, but the spin angular velocity of the fully formed planets was about 30$\%$ smaller than that for the perfect accretion model. They also determined that 49\% of all impact events were hit-and-run collisions, which is also consistent with our estimate. 

In this paper, we focused on the merging criteria for protoplanet collisions. In the future, we plan to investigate additional collisional phenomena such as mantle stripping and ejection of small particles. Using our simulation data for more than 1000 collisional events, we can estimate the change in the core-mantle ratio during the giant impact stage. This is highly relevant to the formation of Mercury, and the formation probability of such planets with very large cores will be investigated.

The ejection of small particles during each collision in the giant impact stage may influence the orbital evolution of terrestrial planets. The ejected material may have damped the eccentricities of the terrestrial planets to their present low values, although it depends on the total amount of material ejected during the giant impact stage. 

\acknowledgments

acknowledgments --- \textit{
We thank Sarah T. Stewart for valuable comments on the manuscript. This research was partially supported by JSPS, the Grant-in-Aid for Young Scientists B (22740291), and MEXT, the Grant-in-Aid for Scientific Research on Priority Areas, and the Special Coordination Fund for Promoting Science and Technology.
}

\clearpage

\clearpage

\clearpage

\clearpage

\clearpage

\clearpage

\clearpage

\clearpage

\begin{table}
\begin{center}
\caption{Parameter sets for the mass ratio of colliding protoplanets\label{mass}}
\begin{tabular}{c||c|c|c}
\tableline\tableline
$\gamma = $ 1 & 0.1$M_\oplus$ vs. 0.1$M_\oplus$ & 0.2$M_\oplus$ vs. 0.2$M_\oplus$ & 0.5$M_\oplus$ vs. 0.5$M_\oplus$ \\
2/3 & 0.2$M_\oplus$ vs. 0.3$M_\oplus$ & --- &  --- \\
1/2 & 0.1$M_\oplus$ vs. 0.2$M_\oplus$ & --- & --- \\
1/3 & 0.1$M_\oplus$ vs. 0.3$M_\oplus$ & --- & --- \\
1/4 & 0.1$M_\oplus$ vs. 0.4$M_\oplus$ & 0.2$M_\oplus$ vs. 0.8$M_\oplus$ & 0.3$M_\oplus$ vs. 1.2$M_\oplus$ \\
1/6 & 0.1$M_\oplus$ vs. 0.6$M_\oplus$ & --- & --- \\
1/9 & 0.05$M_\oplus$ vs. 0.45$M_\oplus$ & 0.1$M_\oplus$ vs. 0.9$M_\oplus$ & 0.2$M_\oplus$ vs. 1.8$M_\oplus$ \\
\tableline
\end{tabular}
\end{center}
\end{table}

\end{document}